\documentstyle[12pt,epsfig]{article}
\begin{document}

\title{Cluster expansions in dilute systems:
applications to satisfiability problems and spin glasses}
\vskip 10pt
\author{
Guilhem Semerjian$^{1}$ and Leticia F. Cugliandolo$^{1,2}$
\\
$^1$Laboratoire de Physique Th{\'e}orique de l'{\'E}cole Normale 
Sup{\'e}rieure, 
\\
24 rue Lhomond, 75231 Paris Cedex 05, France
\\
$^2$Laboratoire de Physique Th{\'e}orique  et Hautes {\'E}nergies, Jussieu, \\ 
1er {\'e}tage,  Tour 16, 4 Place Jussieu, 75252 Paris Cedex 05, France
 \\
{\tt guilhem@lpt.ens.fr, leticia@lpt.ens.fr}
}
\date\today
\maketitle

\begin{abstract}
We develop a systematic cluster expansion for dilute systems in the highly
dilute phase. We first apply it to the calculation of the entropy
of the K-satisfiability problem in the satisfiable phase.
We derive a series expansion in the control parameter, the average
connectivity, 
that is identical to the one obtained by using the
replica approach with a replica symmetric ({\sc rs})
{\it Ansatz}, when the order parameter
is calculated via a perturbative expansion in the control parameter.
As a second application we compute the free-energy of the Viana-Bray model
in the paramagnetic phase.
The cluster expansion 
allows one to compute finite-size corrections in a simple manner and
these are particularly important in optimization problems.
Importantly enough, these calculations prove the exactness of the 
{\sc rs} {\it Ansatz} below the percolation threshold and might require 
its revision between this and the easy-to-hard transition.

\vspace{.5cm}
\noindent LPTENS/0046, LPTHE/0109
\end{abstract}

\section{Introduction}

Very few analytical tools have been successfully employed 
to study disordered systems beyond mean-field. Mainly, one can 
mention the functional renormalization group analysis~\cite{frg},
high-temperature expansions
of finite dimensional
systems~\cite{highT}, expansion in the concentration 
of disordered models defined on finite dimensional
lattices~\cite{largeD} 
and expansions around mean-field 
theories~\cite{cirano}. The replica
method has been used to study dilute spin-glass 
models~\cite{viana}-\cite{Biroli}
and, 
even if it allows one to obtain 
a number of analytical results,
it has revealed particularly 
difficult to implement when applied to dilute systems. It is then 
desirable to develop other analytical methods to treat these 
problems, at least in their simplest phase. In this 
article we investigate an independent analytic approach that is 
based on a cluster 
expansion. It  allows one to compute
several ``additive'' quantities of interest in dilute systems 
such as the energy density, the entropy, the free-energy, etc.
We shall apply this tool to the study of  
two standard problems, with definitions recalled below,  
that are random K-satisfiability (K-sat)~\cite{K-sat} 
and the Viana-Bray dilute 
spin-glass~\cite{viana}. 
The method is similar
to one of the techniques used by Weigt and 
Hartmann~\cite{Weigt} in their study of the vertex-cover problem
on a random graph. The application 
to other dilute systems is straightforward. 
Some of the advantages of this method 
with respect to  replicas are: 
it allows one to compute the corrections to the thermodynamic
limit in a simple way; it allows one to pinpoint a possible limitation of the 
replica symmetric ({\sc rs}) {\it Ansatz} in the satisfiable and paramagnetic 
({\sc pm})  phases of dilute disordered systems
 and, not less importantly, it can be straightforwardly adapted to study 
the evolution of some of the algorithms developed to analyze 
K-sat numerically~\cite{new}. 

The paper is organized as follows. In Section~\ref{K-sat}
we define the random K-sat problem and recall its main 
properties. In Section~\ref{themethod} we define 
the clusters, as well as several useful notions associated to them,
and we introduce the cluster expansion.  Section~\ref{exp} is devoted to
the explicit calculation of the entropy of K-sat.  The finite
$N$ corrections are also described. In Section~\ref{Numer}
we discuss the interplay between the percolation and 
easy-to-hard transition. We
underline the consequences of this
calculation as regards to the validity of the {\sc rs} {\it Ansatz}
in the satisfiable and {\sc pm} phases.
As an application to a physical system, we discuss 
 the calculation of the paramagnetic {\sc pm}
free-energy
of the Viana-Bray~\cite{viana} dilute spin-glass in Section~\ref{vbsection}.
Finally, in Section~\ref{concl} we
present our conclusions.

\section{K-satisfiability}
\label{K-sat}

The theory of complexity has been developed to characterize 
{\it worst}-case instances of 
hard computational problems~\cite{complexity}. 
A classification scheme, according to the time  needed
to find solutions  with the best performing 
algorithms, or to prove that a problem is
not solvable, is one of the outcomes of these studies. 
Of particular importance is the problem 
of K-satisfiability~\cite{K-sat,Goerdt,Cook} (K-sat) that has been 
used as a testing ground for these theories.

However, it has been recently realized
that in many interesting cases in computer science, it is more
relevant to determine the properties of {\it typical}, and not worst, 
realizations of a given  problem~\cite{Fuan}. 
{\it Random} K-sat, defined as 
the ensemble of randomly generated instances of K-sat, is the paradigm
and the goal is now  to predict the behavior of a typical element of 
the ensemble.

The relation between phase transitions, or threshold phenomena, and 
intractability in random combinatorial problems has been
stressed by several authors~\cite{reviews}.
Problems 
that are very hard to solve in the worst case, are not so in the 
typical case, unless the control parameter takes values within 
a finite interval
that defines the critical region. Away from the critical region,
simple algorithms are capable of finding a solution, or showing that
there is no solution, in polynomial time. Random K-sat has a 
well-defined threshold phenomenon.

Random K-sat, as well as other random optimization problems, can be mapped
onto disordered spin models. The mapping is done by associating the
cost function in the optimization problem to an energy density   
in the physical system~\cite{Remi}. Consequently,  
the random character associated to the 
choice of different instances in the optimization problem translates into 
quenched disordered interactions in the physical system. 
The most interesting optimization problems like K-sat
become spin-glass models of a particularly difficult type, 
where each spin interacts with a finite fraction 
of other spins in the sample. These are called dilute spin-glasses and 
they are interesting {\it per se} since they appear as a case of 
intermediate difficulty  between solvable mean-field spin-glasses and 
realistic finite dimensional ones.

The quest of the threshold value of the control
parameter becomes then a search for a phase transition. Thus, 
all tools developed to treat disordered physical systems in
statistical mechanics~\cite{beyond} can be adapted 
to study random optimization problems. In the context of random
K-sat  two main techniques have been used so far: the replica approach 
in the thermodynamic limit~\cite{Remi,Biroli,Remi2,Weigt2}  
and numerical simulations complemented by finite size scaling
when the number of variables remains finite~\cite{scaling1}.
The same two techniques are used in the study of dilute spin-glasses.

Random K-sat is  defined as follows. Consider $N$ Boolean
variables, $\{x_1, \dots, x_N\}$, that can take two logical 
values $x_i=$ TRUE or FALSE, for each $i$. 
Firstly, choose K indices from the set of $N$ elements, $i=1,\dots,N$.
Secondly, assign to each of these indices the {\it literal}
$x_j$, or its negation $\overline x_j$, with equal probability 
$p=1/2$. 
Thirdly, construct a {\it clause} $C_1$ as the logical OR ($\vee$) of the 
K previously determined literals. If K$=3$ and $N=10$ a possible clause is 
$x_1 \vee \overline{x_5} \vee x_7$. 
New clauses are generated in identical manner, independently of the 
previous ones. One usually calls $M$ the total number of clauses. 
A {\it formula}
$F$ is the logical AND ($\wedge$) of $M$ such clauses. It reads
\begin{equation}
\displaystyle{
F=\large{\bigwedge}_{l=1}^M C_l = \bigwedge_{l=1}^M \left(\bigvee_{i=1}^K z_i^l\right)
}
\; ,
\end{equation}
where $z^l_i \in \{x_1, \overline x_1, ... \, , x_N, \overline x_N \}$.
A solution, if it exists, is an      
assignment of the $N$ variables that satisfies $F$, that is to
say, for which all clauses are verified simultaneously. 

Note that in the process of generation of a clause, two 
random processes intervene. In the first one, one selects the 
variables, in the second one, once the variables have been chosen, 
one determines the  requirements that will be imposed on them. 
We shall later take advantage of this two-step process to perform the 
average over disorder in a convenient order.

It is clear that if $M\ll N$ it will be very easy to find a solution
to $F$. On the contrary, if $M\gg N$, it will be extremely difficult to
satisfy all requirements simultaneously. Indeed, a well-defined 
critical value $\alpha_c(\mbox{K})$ of the parameter 
$\alpha\equiv M/N$ appears when $M\to\infty$ and $N\to\infty$ with their ratio
$\alpha$ kept fixed. This limit corresponds to a 
thermodynamic limit, in the physical
language. A  threshold phenomenon, reminiscent of a phase
transition, is observed: for $\alpha < \alpha_c(\mbox{K})$ 
all formul{\ae} have at least
one solution with probability one, whereas for 
$\alpha > \alpha_c(\mbox{K})$ any formula has no solution with
probability one.

Different values of K lead to different critical behaviors. When 
K $=1$, the model is unsatisfiable for all finite values of $\alpha$, 
{\it i.e.} $\alpha_c(\mbox{K}=1)=0$. When K $=2$, K-sat has a continuous phase 
transition at $\alpha_c(\mbox{K}=2)=1$. This is a rigorous result proven   
by using a mapping
on a directed graph problem~\cite{Goerdt}. For K $\ge3$ only numerical
estimates for $\alpha_c(\mbox{K}\geq 3)$ and approximate results obtained with 
the replica method are available~\cite{Remi}, these yield 
$\alpha_c(\mbox{K} = 3)\sim 4.2$. 

The replica method  is a powerful tool of statistical mechanics that 
allows one to compute the statistical properties of a disordered
physical problem  
in equilibrium with a thermal environment. In order to 
use it to study optimization problems in general, and K-sat in particular, 
one first maps the optimization problem 
onto a statistical mechanics model. In the case of K-sat, the physical model
is a spin-glass model with 
dilute interactions of random sign. Indeed, 
a natural representation of any K-sat formula is obtained by introducing 
an $M \times  N$ matrix $C_{li}$, whose elements are
\begin{eqnarray}
C_{li} &=& 
\left\{
\begin{array}{rcr}
0 & \;\;\; \mbox{if} & \;\;\; \mbox{neither} \;  x_i  \; \mbox{nor} \;
\overline x_i \; \in C_l \; , 
\\
1 & \;\;\; \mbox{if} &  \;\;\; x_i \in C_l \;,
\\
-1& \;\;\; \mbox{if} &  \;\;\; \overline{x}_i \in C_l
\; .
\end{array}
\right.
\end{eqnarray}
The random generation of clauses
is equivalent to a uniform distribution
of the matrices $C_{li}$ that satisfy the
constraints $\sum_i C_{li}^2 =$K, $\forall l$.

A cost function for K-sat is given by the number of unsatisfied 
clauses in a given formula. If one identifies the logical state 
$x_i=$ TRUE with a spin $S_i=1$ and the  logical state 
$x_i=$ FALSE with a spin $S_i=-1$, it is then 
easy to verify that the following 
expression counts the number of unsatisfied clauses
\begin{equation}
E[\{C_{li},x_i\}] = \sum_{l=1}^M 
\delta^{(\mbox{K})}\left(\sum_{i=1}^N C_{li} S_i,-\mbox{K} \right) 
\; ,
\end{equation}
where $\delta^{(\mbox{K})}(a,b)$ is the Kronecker delta function.
Using a polynomial representation of $\delta^{(\mbox{K})}$ this expression 
can be rewritten as the total 
energy of a sum of dilute $p$ spin-glass models in a random field
(several values of $p$ intervene, how many depends 
on the value of $K$)~\cite{Remi}.  

Once the energy function is identified, 
one introduces a fictive temperature 
$T$, then computes the average free-energy 
with the help of the replica trick, 
and finally takes the limit $T\to 0$ to study the ground state 
properties of the physical model. This gives access to quantities such 
as, for example, the average entropy of the satisfiable phase. 
This is defined as the average over disorder of the logarithm of the 
number of solutions.
One of the drawbacks of the use of the replica method is that an
{\it Ansatz} is necessary to pursue the calculation. 
Even in the simplest phases, the 
satisfiable one for K-sat, it is not obvious to show that the
simplest {\it Ansatz}, called replica symmetric ({\sc rs}), 
solves the problem exactly.
Moreover, it has been proven that 
 in the unsatisfiable phase one has to go beyond the 
{\sc rs} {\it Ansatz} and develop a replica symmetry breaking ({\sc
rsb})  scheme. This is indeed 
a very difficult task since 
the order parameters for dilute systems have a much more
intricate structure than for infinitely connected
cases~\cite{viana,diluted,other_diluted,KS}. Recent 
progress in this direction has been presented in Ref.~\cite{Biroli}.


In this paper we re-derive a generic expression for the entropy of
K-sat using a very simple method
that avoids the use of replicas. Furthermore, the method allows us to 
compute the  finite-size corrections. Our derivation gives information
about the domain of validity of 
Monasson and Zecchina's conjecture that the {\sc rs} {\it Ansatz} 
is exact in the satisfiable phase.
We explain the expansion using the formalism of 
K-sat but the line of reasoning can be applied to any dilute system
in the dilute regime. In Section~\ref{vbsection} we shall 
analyze the Viana-Bray model~\cite{viana} along the same lines.
 
\section{The method}
\label{themethod}

Let us start this Section 
by setting the notation and defining a set of notions that 
will be used later.

Given a formula $F$ of K-sat, two variables $x_i$ and $x_j$ are called
{\it adjacent} if there is at least one clause in $F$ in which both
$x_i$ and $x_j$ appear, irrespectively of the fact that they are
negated or not.
Two variables are {\it connected} if and only if 
there is a path of adjacent variables
between them. A {\it cluster} is a 
set of connected variables that are disconnected from all others. 
Let us label with an integer $r$ the different clusters of the formula $F$,
$r=1,\dots,{\cal N}_c(F)$, where ${\cal N}_c(F)$ is the 
total number of clusters in $F$. We shall call 
$n_0(F)$ the number of variables that do not belong to any cluster. 

\begin{figure}
\begin{center}
\epsfig{file=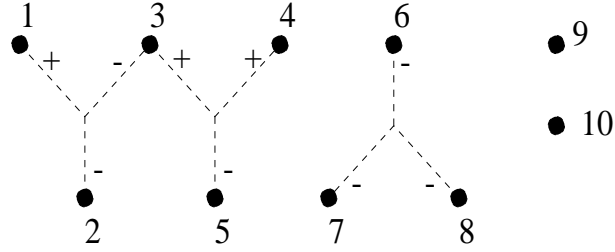,width=8cm,height=3.2cm}
\vspace{.25cm}
\caption{A graphical representation of the formula $F=(x_1 \vee
\overline{x_2} \vee \overline{x_3}) \wedge (x_3 \vee x_4 \vee
\overline{x_5}) \wedge (\overline{x_6} \vee \overline{x_7} \vee
\overline{x_8})$. See the text for details.}
\label{ex1}
\end{center}
\end{figure}

These definitions are very easy to picture. For instance, take a 
3-sat problem with ten variables, 
$i=1, \dots, 10$, that is  defined by the formula 
$F=(x_1 \vee
\overline{x_2} \vee \overline{x_3}) \wedge (x_3 \vee x_4 \vee
\overline{x_5}) \wedge (\overline{x_6} \vee \overline{x_7} \vee
\overline{x_8})$. The variables $i=9$ and $i=10$ do not belong to 
any cluster, thus $n_0(F)=2$. A graphical representation of each 
clause is very useful. We associate a point to each variable.
Each clause is represented by a 
{\it star} with K legs, $3$ in the example,  
with endpoints that represent the variables. 
In the formula $F$ there are two clusters, ${\cal N}_c=2$, 
that link $i=1, \dots,5$ and $i=6,7,8$, respectively. 
When a variable appears in two (or more) clauses it will be 
shared by two stars. This is the case in the cluster on the left of 
Fig.~\ref{ex1}. More complicated structures are possible, particularly
when $N$ and $M$ are large. 
The assignment $x_i$ ($\overline x_i$) 
of the $i$th literal in a clause can be represented
with a plus (minus) sign on its leg.  
These are the  signs in Fig.~\ref{ex1}.
In this way, a one-to-one 
correspondence between formul{\ae} and graphs is constructed.

When $\alpha$ is small, the typical cluster size is expected to be
small as well, as there are much less clauses than variables. Indeed,
for K $=2$ this problem is the one of percolation in an infinite
dimensional space, also known as the random graph. Many properties
of this model are known~\cite{Bollobas}, among which the fact that
for $\alpha<1/2$ all variables belong to clusters of size at most 
proportional to $\ln N$,  in the
thermodynamic limit. When $\alpha$ crosses $1/2$ 
a giant cluster
containing a finite fraction of the variables grows continuously. For
K $\ge 3$ the equivalent geometrical problem relies on the theory of
hyper-graphs, for which less is known. The percolation occurs at
$\alpha=\alpha_p \equiv 1/(\mbox{K}(\mbox{K}-1))$~\cite{hypergraphs}. 
In these two
cases the system percolates much before it becomes
unsatisfiable, $\alpha_p < \alpha_c$. Indeed the appearance of a 
contradiction requires an
intricate structure in the giant cluster.

Let us define the ground state entropy of a  formula
$S_{\sc gs}(F)$ as the logarithm of the number of assignments of the
variables that minimize the number of violated clauses. If $F$ is
satisfiable, $S_{\sc gs}(F)$ is the logarithm of the 
number of solutions of $F$. It is
clear from the clusters' definition that $S_{\sc gs}$ is the
sum of contributions of the different independent sub-formul{\ae}:
\begin{equation}
S_{\sc gs}(F)=n_0(F) \ln 2 + 
\sum_{r=1}^{{\cal N}_c(F)} S_{\sc gs}(F_r)
\; .
\label{sum}
\end{equation}

We are interested in the entropy averaged over
the ensemble of formul{\ae}, $\overline{S_{\sc gs}}$. 
We shall henceforth  denote ensemble averages with an over-bar.
As stressed in Section~\ref{K-sat} this average is twofold.
Indeed, clusters can be separated into ensembles with the same topology,
ignoring for the moment the sign assignment of the literals.
Thus, the averaging proceeds in two steps; one  
first chooses the topology of the cluster, with its associated 
probability, and then one averages over the 
two possibilities for each literal in the cluster. 
For a given cluster, once the
latter average is performed, the entropy depends only on the topology
of the cluster. This remark allows us to rewrite the average of the 
sum in  Eq.~(\ref{sum}) in a more convenient manner.
If we introduce a new integer  $t$ 
that labels all possible topologies, and 
$n_t(F)$ and $\langle S_t\rangle $ the number of $t$-like clusters 
in formula $F$ and the average over the sign assignement of the entropy
of the $t$-like clusters, respectively, we arrive at the following
expression for the averaged entropy:
\begin{equation}
\overline{S_{\sc gs}}= \sum_t [n_t] \langle S_t\rangle
\; .
\end{equation}
We have here included the isolated variables in the sum, associating them 
to the index $t=0$, $S_0=\ln 2$ and we denote with $[n_t]$ the
average number of $t$-like clusters.

A more convenient expression for $[n_t]$ can now be worked out. 
Let us call $X_t^i(F)$ the function which takes the value  
$1$ if the variable $i$ belongs to
a $t$-like cluster of the formula $F$ and $0$ otherwise; let 
$L_t$ be the number of variables in such a cluster. Then
\begin{equation}
n_t(F)=\frac{1}{L_t} \sum_i X_t^i(F)
\; ,
\end{equation}
that implies 
\begin{equation}
\frac{1}{N} \, [n_t] = 
\frac{1}{L_t} [X_t^1] = \frac{P_t}{L_t}
\; ,
\end{equation}
where $P_t\equiv [X_t^1]$ is the probability that a given variable
belong to a $t$-like cluster. Finally, 
\begin{equation}
\frac{1}{N} \, \overline{S_{\sc gs}}=\sum_t \frac{1}{L_t}P_t 
\langle S_t\rangle
\; .
\label{new_eq}
\end{equation}
$P_t$ and $\langle S_t \rangle$ 
can now be obtained using elementary combinatorial 
arguments and simple enumeration. 

This formulation can be adapted to any quantity for which the clusters
contribute additively, the free-energy for instance, and to other
dilute problems where there is also a decoupling in the randomness
between a geometrical part and an interaction one, as in the
Viana-Bray model~\cite{viana}.  

\section{Cluster expansion of the K-SAT entropy}
\label{exp}

In this Section we shall apply the cluster expansion
to the calculation of the average entropy of 
random K-sat. For our present purposes K$=1$-sat is not 
interesting since it is unsatisfiable for all finite 
values of $\alpha$. We shall then 
start by analyzing in detail K$=2$-sat. Afterwards, we shall 
discuss how the approach generalizes to larger values of K.

\subsection{K${\bf=2}$-sat in the thermodynamic limit}

For a cluster of $n$ variables connected by $p$ distinct 
clauses, the probability $P_t$ reads
\begin{eqnarray}
P_t &=& 
p!
\left(
\begin{array}{c}
M \\ p
\end{array}
\right)
\left(\frac{2}{N(N-1)}\right)^{p}\left(\frac{(N-n)(N-n-1)}{N(N-1)}
\right)^{M-p}
\nonumber
\\
& & \times  \;
(n-1)! 
\left(
\begin{array}{c}
N-1 \\ n-1
\end{array}
\right) \, 
K_t \; . 
\label{Pt}
\end{eqnarray}
Let us briefly describe the origin of the  factors in this equation.
Each of the $p$ clauses is chosen with probability $2/(N(N-1))$
at each of the $M$ steps in the formula generation process. For the 
variables belonging to the cluster to be disconnected from all other 
sites, the $M-p$ other clauses must belong to the set of the
$(N-n)(N-n-1)/2$ clauses connecting the other sites. The first
two factors come from the possible permutation of the 
$p$ steps where the considered clauses appear. The last three
factors arise from the freedom in the choice of $n-1$ sites 
connected to the chosen site. 
In particular, $K_t$ is a symmetry factor that equals 
 the number of distinct labellings of the $n$ sites, divided
by $(n-1)!$. Note that two labellings which lead to the same set of
clauses are {\em not} distinct: for the linear three site cluster
$1-2-3$ and $3-2-1$ correspond to the same labelling, with clauses
$(12)$ and $(23)$. But $1-2-3$ and $2-1-3$ are distinct.

\begin{figure}[t]
\begin{center}
\epsfig{file=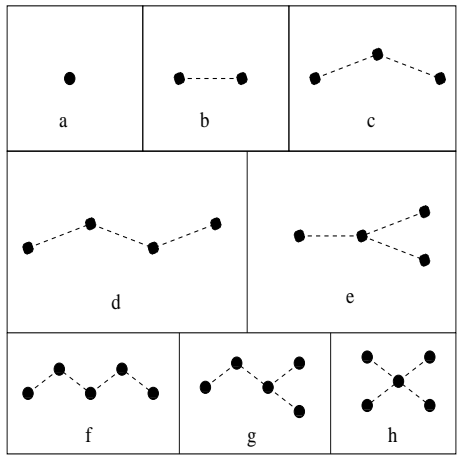,width=6cm,height=6cm}
\vspace{.25cm}
\caption{Tree-like clusters that contribute to K $=2$-sat.}
\label{clusters_K2}
\end{center}
\end{figure}

In the thermodynamic limit $N\to\infty$ and  $\alpha$ fixed, 
and for $n$ and $p$ finite, this expression
is proportional to $N^{n-1-p}$ ($p \ge n-1$). 
It is then finite only if $p=n-1$, 
that is to say for tree-like clusters. This justifies the choice of 
distinct clauses.
In this limit, for $p=L_t-1$ and $n=L_t$, 
this expression simplifies greatly:
\begin{equation}
P_t=(2\alpha)^{L_t-1}e^{-2 L_t \alpha}K_t
\; .
\end{equation}
The different clusters considered in the expansion are represented in 
Fig.~\ref{clusters_K2}.
For each type, the relevant quantities, obtained by basic enumeration,
are given in Table~\ref{table1}. 
For instance, the average entropy of the linear three sites cluster
is made of two parts: if the clauses require the same sign for the
central variable, one can find five solutions of the formula; if
the clauses are contradictory for the central variable, there are only
four solutions.  

\begin{table}
\vspace{-.5cm}
\begin{center}
\begin{tabular}{||c||c|c|c||}
\hline
\hline
$\mbox{type}$ & $L_t$ & $ K_t$ & $\langle S_t \rangle $ 
\\ 
\hline
a & $1$ & $1$ & $\ln 2$
\\
\hline
 b & $2$  & $1$ & $\ln 3$
\\
\hline
c & $3$ & $3/2$ & $(2\ln 2 + \ln 5)/2$
\\ 
\hline
d & $4$ & $2$ & $(3 \ln 2 + \ln 5 + 2 \ln 7)/4$
\\
\hline
e & $4$ & $2/3$ & $(3 \ln 2 + 5\ln 3)/4$
\\
\hline
f & $5$ & $5/2$ & $(4\ln 2 +6\ln 3 +\ln 5 +2 \ln 11 + \ln 13)/8$
\\
\hline
g & $5$ & $5/2$ & $(9\ln 2 + 2 \ln 5 + 2 \ln 7 + \ln 11 + \ln 13)/8$
\\
\hline
h  & $5$ & $5/24$ & $(13 \ln 2 + 4 \ln 5 + \ln 17)/8$
\\
\hline
\hline
\end{tabular}
\vspace{.25cm}
\caption{The contributions of the clusters in Fig.~\ref{clusters_K2}}
\label{table1}
\end{center}
\end{table}


Expanding in $\alpha$ up to $O(\alpha^4)$ we obtain
\begin{eqnarray}
\frac{1}{N} \overline{S_{GS}} = 
\;\;\;\;\;\;\;\;\;\;\;\;\;\;\;\;\;\;\;\;\;\;\;\;\;\;\;\;\;\;\;\;\;\;\;\;\;
\;\;\;\;\;\;\;\;\;\;\;\;\;\;\;\;\;\;\;\;\;\;\;\;\;\;\;\;\;\;\;\;\;\;\;\;\;\;\;
\;\;\;\;\;\;\;\;\;\;\;\;\;\;\;\;\;\;\;\;\;\;\;\;\;\;\;\;\;\;\;\;
& & 
\nonumber\\
\ln 2 + \alpha \ln \left(\frac{3}{4}
\right) + \alpha^2 \ln \left(\frac{80}{81}\right) + 
\frac{\alpha ^3}{3} \ln \left(\frac{3^{29} 7^6}{5^{15}  2^{28}}\right) +
\frac{\alpha^4}{12} \ln \left( \frac{2^{225}5^{160}11^{36}13^{24}17}
{3^{216}7^{168}}\right) 
\; .
& & 
\nonumber
\\
\label{exp_Ninfty}
\end{eqnarray} 
Monasson and Zecchina obtained this series by using the 
replica trick, with a {\sc rs} {\it Ansatz},
to average the free-energy of the physical model 
related to 2-sat~\cite{Remi}. The averaged entropy follows from the averaged 
free-energy density that itself depends on the probability 
distribution of the local fields. 
This quantity is determined by an integro-differential
equation that cannot be solved analytically. Monasson and 
Zecchina  developed a 
perturbative solution in $\alpha$ that allowed them to derive a series
for $\overline S_{\sc sg}/N$ that coincides, up to $O(\alpha^4)$, with the
one in Eq.~(\ref{exp_Ninfty}). The perturbative nature of this 
result is now clarified from the cluster analysis. 
Note that we performed two expansions to obtain this series: 
the cluster enumeration and an expansion in powers of $\alpha$ 
of the exponentials in $P_t$.
We shall further 
discuss this issue in Section~\ref{Numer}.

\subsection{Finite size corrections to the entropy of K{\bf$=2$}-sat}

There are two kinds of finite size corrections to the expansion
presented in Eq.~(\ref{exp_Ninfty}). On the one hand,  
the probability $P_t$ of a variable belonging to a tree-like cluster has
$1/N$ corrections that can be simply computed from the 
general expression~(\ref{Pt}). On the other hand, 
clusters that include loops also contribute to the $1/N$ corrections.

The expansion of expression (\ref{Pt}) up to order 
$1/N$ for tree-like clusters with $n=L_t$ and $p=L_t-1$ yields
\begin{eqnarray}
P_t&=&(2 \alpha)^{L_t-1} e^{-2L_t \alpha} K_t 
\nonumber
\\ & & \times 
\; \left(1+\frac{1}{N}
\left[2L_t(L_t-1)-\alpha L_t(L_t+1) -
\frac{(L_t-1)(L_t-2)}{2}\left(1+\frac{1}{\alpha}\right)\right]\right)
\; .
\nonumber\\
\label{nueva}
\end{eqnarray}

\begin{figure}[h]
\begin{center}
\epsfig{file=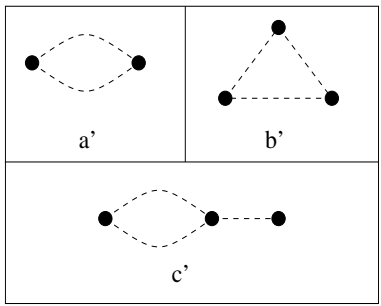,width=5cm,height=4cm}
\end{center}
\caption{Loop diagrams contributing to the $1/N$ corrections.}
\label{loops}
\end{figure}

Clusters with $l$ loops contribute to the order $1/N^l$. Hence, if we 
only wish to compute the $1/N$ corrections  we can
content ourselves with clusters that have only one loop. These 
have $L_t$ variables and also $L_t$ clauses. One obtains
\begin{equation}
P_t=\frac{1}{N} (2\alpha)^{L_t} e^{-2L_t \alpha} K_t
\; ,
\label{Pt1N}
\end{equation}
with $K_t$ defined as before and multiplied by $1/2$ if 
there is a repeated clause.
The one-loop clusters that we considered are represented in
Fig.~\ref{loops}.

\begin{table}[h]
\begin{center}
\begin{tabular}{||c||c|c|c||}
\hline\hline
$\mbox{type}$ & $L_t$ & $K_t$ & $\langle S_t \rangle $
\\
\hline
a' & $2$ & $1/2$ & $(3\ln 2 +\ln 3)/4$
\\
\hline
b' & $3$ & $1/2$ & $(9\ln 2 + 3\ln 3)/8$
\\
\hline
c' & $3$ & $3/2$ & $(5 \ln 2 +4\ln 3 +\ln 5)/8$
\\
\hline
\hline
\end{tabular}
\vspace{.25cm}
\caption{The contributions of the clusters represented in Fig.~\ref{loops}.}
\label{table2}
\end{center}
\end{table}

Including the $1/N$ corrections in Eq.~(\ref{nueva}) 
and the ones stemming from the new diagrams in Fig.~\ref{loops} and 
Eq.~(\ref{Pt1N}) 
calculated with the results of Table~\ref{table2}, the correction 
to $\overline{S_{\sc gs}}/N$ reads
\begin{equation}
\frac{1}{N}\left[\alpha\ln\left( \frac{3^4}{2^4 5}\right) +
\frac{\alpha^2}{4}\ln \left(\frac{2^{107}5^{56}}{3^{107}7^{24}}\right) 
+\frac{\alpha^3}{2} \ln\left(\frac{3^{193}7^{156}}{2^{199}5^{141}11^{36}13^{24}17}\right)
\right]
\; .
\end{equation}
This result has to be checked against exhaustive numerical evaluation of
the entropy for small systems.

\subsection{K $\mathbf{\ge 3}$-sat in the thermodynamic limit}

The method described in detail for K $=2$ 
can be used for any value of K. As the graphical representation
and the enumeration of clusters are, however, 
more cumbersome than for  K $=2$ 
we shall present less detailed results for the case K $\geq 3$.

\begin{figure}[h]
\begin{center}
\epsfig{file=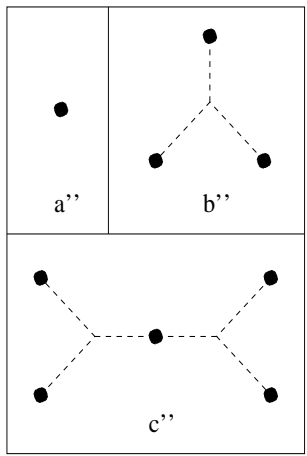,width=4cm,height=6cm}
\vspace{.25cm}
\caption{Clusters yielding the leading 
contributions to K $=3$-sat.}
\label{k3_clusters}
\end{center}
\end{figure}

The probability for a given variable to be present in a cluster of
$L_t$ variables that are linked by $p$ clauses is of order $1$ in the
thermodynamic limit only if
$p(K-1)=L_t-1$, which is the tree-like condition for these
hyper-graphs. If this holds
\begin{equation}
P_t=(\alpha K!)^p e^{-L_t \alpha K} K_t
\; .
\end{equation}
In Fig.~\ref{k3_clusters} we have drawn
the  diagrams leading to the main contributions 
for K $=3$. In the text we give the analytic expression
for  general K.

\begin{table}[h]
\begin{center}
\begin{tabular}{||c||c|c|c||}
\hline
\hline
type & $L_t$ & $K_t$ & $\langle S_t \rangle $ 
\\
\hline
a'' & $1$ & $1$ & $\ln 2$
\\
\hline
b'' & $K$ & $K/K!$ & $\ln (2^K-1)$
\\
\hline
c'' & $2K-1$ & $K^2(2K-1)/(2(K!)^2)$
&
$\frac{1}{2}\ln(1+2^K(2^{K-1}-1)) + \frac{1}{2}\ln (2^K(2^{K-1}-1))$
\\
\hline
\hline
\end{tabular}
\end{center}
\caption{The contributions of the clusters represented in 
Fig.~\ref{k3_clusters}.}
\end{table}
With these values we obtain
\begin{eqnarray}
\frac{1}{N}\overline{S_{GS}}
&=&
\ln 2 + \alpha \ln \left(1-\frac{1}{2^K}\right)
\nonumber\\
& &
\hspace{-2.2cm}
+\frac{\alpha^2 K^2}{2} \left[ - \ln\left(1-\frac{1}{2^K}\right) 
+\frac{1}{2} \ln \left( 1-\frac{1}{2^K-1}\right)
+\frac{1}{2} \ln 
\left( 1-\frac{2^{K-1}-1}{2^{K-1}(2^K-1)}\right)\right]
\; .
\label{exp_Ninfty_K3}
\end{eqnarray}
Again we recover the {\sc rs} result of Ref.~\cite{Remi}.
The contributions to the finite size corrections are similar to the ones 
discussed for 2-sat; we obtain at the leading order in $\alpha$:
\begin{equation}
\frac{1}{N} \frac{\alpha K^2}{2} \left[ \ln\left(1-\frac{1}{2^K}\right) 
-\frac{1}{2} \ln \left( 1-\frac{1}{2^K-1}\right)
-\frac{1}{2} \ln 
\left( 1-\frac{2^{K-1}-1}{2^{K-1}(2^K-1)}\right)\right]
\; .
\end{equation}

We have shown that the perturbative analysis of the 
{\sc rs} {\it Ansatz} leads to a series
in $\alpha$ which leading orders  coincide with the ones 
stemming from the {\it finite} cluster expansion, for all K. 
In the next Section 
we shall discuss the relevance  of the 
contributions from the infinite cluster that appears at the percolation 
transition $\alpha_p < \alpha_c$.

\section{Discussion}
\label{Numer}

The domain of validity of our expansion, and of the 
results obtained with the replica method~\cite{Remi},
can be enlighted by studying the percolation phenomenon
in detail.

The series expansion in (\ref{new_eq}) is ordered following the index $t$ 
that is directly related to the size of the clusters. The dependence
on $\alpha$ is involved since the coefficients $P_t$ are
$\alpha$-dependent via an exponential times a power.
The expansion of the exponential factors in powers of $\alpha$ leads
to a rearrangement of the series in powers of $\alpha$.

The range of validity of both series is not obvious. We can start 
by analysing the simpler series  $\sum_t P_t$
that should count the total fraction of sites and be identical to 
$1$. For K$=2$, its direct summation 
yields $1$ for $\alpha<\alpha_p$ and  $1-P$ for $\alpha>\alpha_p$,
where $P$ is the solution to 
$1-P=e^{-2\alpha P}$ (see  Eq.~(\ref{eq_P}) below).
Thus, it fails above $\alpha_p$ because of the emergence of a giant
cluster at the percolation transition. Instead, if one expands the
exponentials in powers of $\alpha$, 
the result is $\sum_t P_t=1 + 0 \times \alpha + 0\times \alpha^2
+\dots$. The sum yields $1$ independently of $\alpha$ even beyond the  
percolation threshold. The rearragement in powers of $\alpha$ captures
the correct behavior of this quantity.

One can conjecture that 
the rearrangement yields the exact result for all
$\alpha$ for all  quantities that depend mainly on the locally 
tree-like structure of the percolating cluster, and only weakly on its
loops, which 
show up only on a scale of order $\ln N$. As can be seen in Eq.~(9),
the exponentials 
in $P_t$ arise from the requirement that the considered cluster is
disconnected 
from the rest of the sites. Expanding the exponentials amounts
basically to assuming
that a sub-tree of the giant cluster gives the same contribution as
its disconnected 
counterpart. The alternative signs arise from the need to avoid the
 double counting of the smaller clusters contained in the giant cluster.

As regards to the calculation of the averaged entropy, 
if one could compute all the terms in (\ref{new_eq}) 
and sum the series, the result would be exact under the percolation
threshold, $\alpha_p=1/(K(K-1))$. Indeed all 
sites belong to clusters of size at most proportional to $\ln N$ in
this regime. 
As soon as $\alpha$ goes beyond $\alpha_p$, the direct 
summation of (\ref{new_eq}) should fail.

In spite of the discussion of the next to last paragraph, 
we do not expect that reordering the series in powers of $\alpha$ 
gives the correct result for the entropy. 
Our argument is based on the drastic influence of loops on 
this quantity. Let us compare the entropy of a loop and of a linear 
cluster of the same size, for $2$-sat. 
There are roughly twice as many solutions for
the linear 
cluster as for the loop, as one does not require the ending
variables to be 
the same. The difference of entropy between the two should then be
finite. As 
there are an extensive number of loops in the percolating cluster, one can expect a
 finite deviation in the average entropy per site between the result
assuming 
a tree-like structure of the giant cluster (i.e. the expansion in
alpha of the 
original series) and the correct one.

We have examined these issues with the help of 
numerical simulations of systems with small sizes.

\begin{figure}
\begin{center}
\epsfig{file=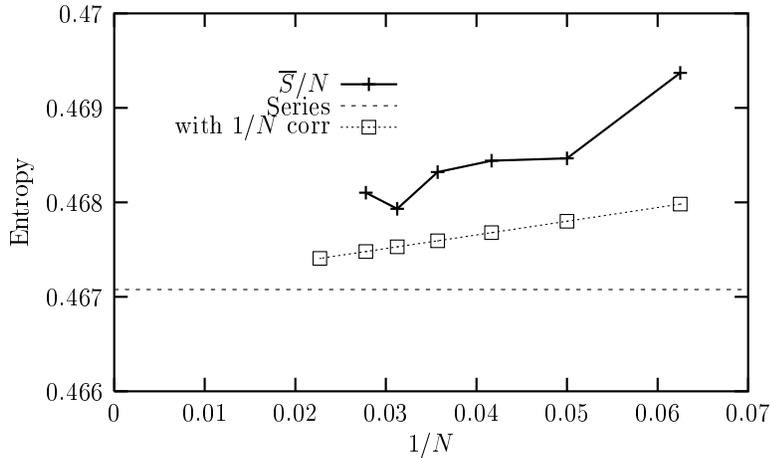,width=12cm}
\vspace{.25cm}
\caption{The averaged entropy per site
for $K=2$ and $\alpha=0.75$. The sizes are $N=16,20,24,28,32,36$.
The lower straight line indicates the
value given by the truncated series expansion found with the replica 
method in the thermodynamic limit (upto $O(\alpha^8)$) 
and the upper curve with linepoints
includes the $1/N$ corrections, for each value of $N$.
}
\label{entropy_tot}
\end{center}
\end{figure}

Firstly, we compare the averaged   total entropy per degree 
of freedom, $\overline{S}/N$, to the 
value predicted by the series expansion once reordered in powers
of $\alpha$. 
In Fig.~\ref{entropy_tot} we plot  $\overline{S}/N$
against $1/N$ with linespoints ${+}$ for K=2-sat problems with $N=16 (50000), 
20 (50000)$, $24 (50000)$, 
$28(10000)$,
$32 (15000)$, $36 (10000)$ and 
$\alpha_p< \alpha=0.75<\alpha_c$. 
For each sample we computed the entropy by exhaustive enumeration.
The numbers between parenthesis are the number of realizations
of random instances of K-sat used to compute the averages.
The accord with the analytical prediction of the truncated series
expansion in the thermodynamic limit (horizontal line below) and 
including $1/N$ corrections (tilted line above)
is very good within $0.3\%$.
We have also computed the variance, 
$1/N \, (\overline{S^2} -{\overline{S}}^2)$, and checked that it is in 
good accord with the analytical 
result, $0.025 \alpha^2$, that we 
obtained with an extension of the cluster expansion described 
in previous sections.

\begin{figure}
\begin{center}
\epsfig{file=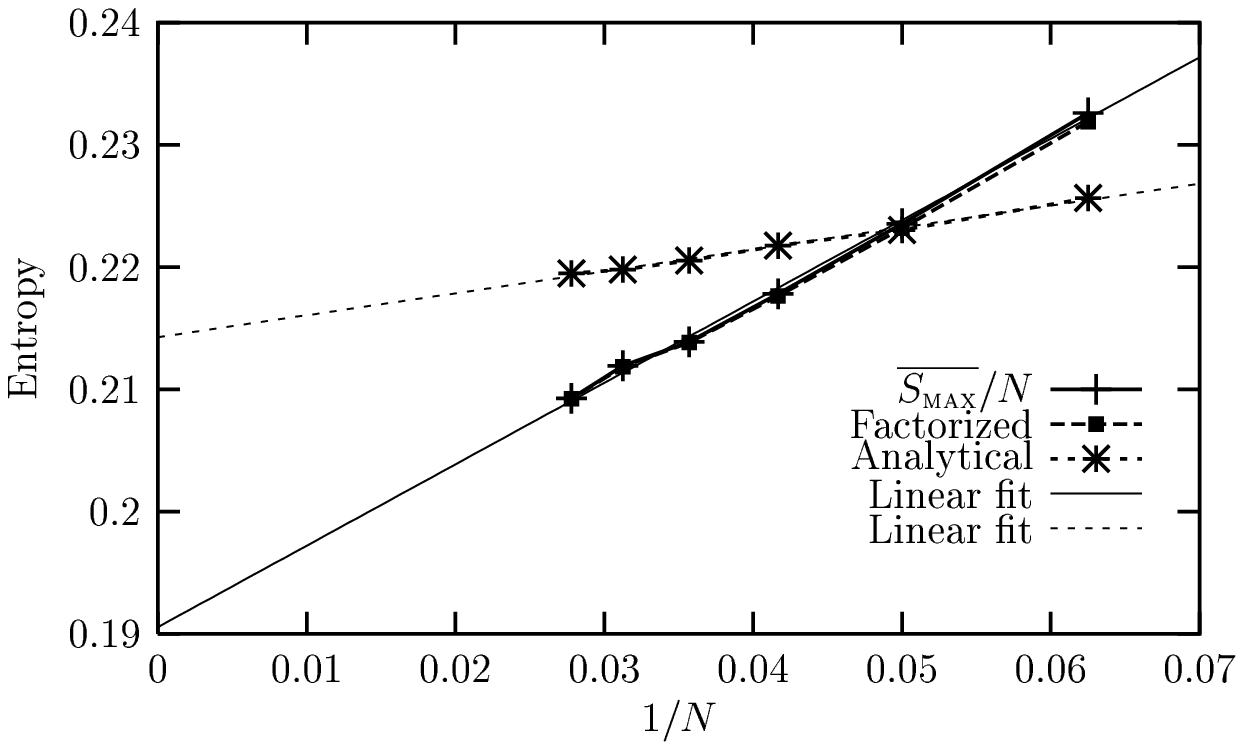,width=12cm}
\vspace{.25cm}
\caption{The averaged entropy of the maximum cluster 
per variable, $\overline{S_{\sc max}}/N$,
the factorized average
$\overline{S_{\sc max}/L_{\sc max}} \; \overline{L_{\sc max}}/N$
and the semi-analytical prediction $P_{\sc max}  
\overline{S_{\sc max}/L_{\sc max}}$ as a function of $1/N$ 
for $\alpha=0.75$ and $K=2$.}
\label{entropy_max}
\vspace{.35cm}
\epsfig{file=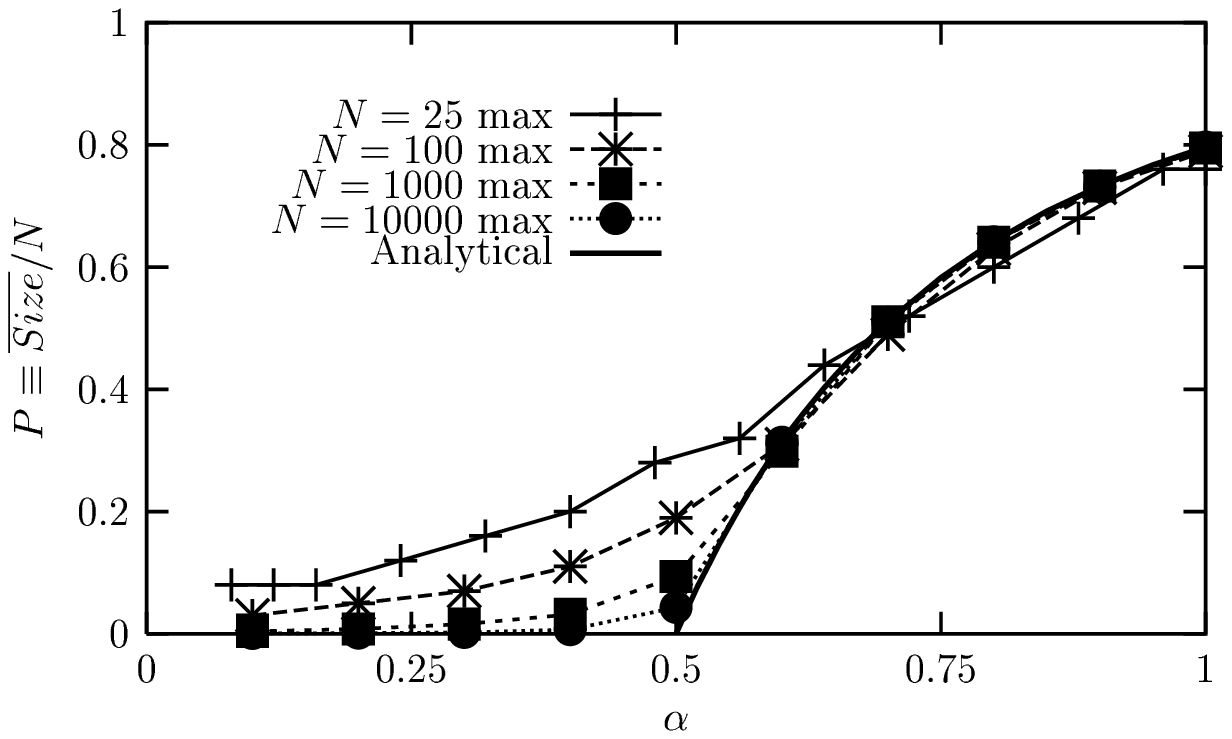,width=12cm}
\vspace{.25cm}
\caption{$K=2$. The fraction of sites belonging to the
largest cluster.
The asymptotic limit $N \to\infty$ is approximately reached
for $N\sim 10000$ where $\overline{L_{\sc max}}/N$ is ``finite'' only above the
percolation threshold $\alpha_p(K=2)=0.5$. The analytical expression
is given by the solution to Eq.~(\ref{eq_P}) and it fits the data very
accurately for $\alpha > 0.6$ as soon as $N>1000$. We have verified
that all other variables are distributed among finite clusters.}
\label{k2clusters}
\end{center}
\end{figure}

Even if the accord between numerical results and theory is 
almost perfect for these small sizes, 
a more careful inspection of the different contributions to the 
total entropy shows that one could expect important deviations 
for increasing $N$. We have computed separately 
the contribution of the largest cluster. Figure~\ref{entropy_max}
represents its study. We here plot, with 
crosses, the averaged entropy of the largest 
cluster, per degree of freedom, $\overline{S_{\sc max}}/N$. 
Its contribution, even for these small sizes, represents 
roughly a half of the total entropy (note that below the 
percolation threshold the contribution of the largest 
cluster is much smaller) and 
that does not seem to vanish in the thermodynamic 
limit. A simple linear fit of 
$\overline{S_{\sc max}}/N$ yields, when $N\to \infty$, the finite limit
$0.1906\pm 0.0008$. This may be taken as a guess for a 
lower limit for the 
contribution of the largest cluster.

In the same figure we compare 
the averaged entropy of the largest cluster   
$\overline{S_{\sc max}}/N$ to the ``factorized'' quantity 
$\overline{L_{\sc max}}/N \; 
\overline{S_{\sc max}/L_{\sc max}}$, where $L_{\sc max}$
is the number of variables in the largest cluster, that is represented
with squares.
For finite sizes we have shown that these two quantities 
coincide. The good accord between the two curves 
suggests that the factorization 
also holds when $N\to\infty$. 

This observation suggests an improvement of the 
finite-size numerical study.
 If we {\it assume} that the factorization  
holds in the limit $N\to \infty$ for the percolating cluster, 
we can then replace the fraction of sites 
that belong to the largest cluster,
$P_{\sc max} \equiv \overline{L_{\sc max}}/N$, 
by its analytical value in the thermodynamic limit. This is given by 
\begin{equation}
1- P_{\sc max} = 
\exp\left(\alpha K \left( (1-P_{\sc max})^{K-1}-1\right) \right)
\; .
\label{eq_P}
\end{equation}
This result is obtained using  a self-consistent equation on the 
generating function that counts the number of sites in finite size 
clusters~\cite{Bray-Rodgers}. 
For $K=2$ and $\alpha=0.75$, 
$P_{\sc max}\sim 0.5828$.
Figure~\ref{k2clusters} displays $P_{\sc max}$ as a function of 
$\alpha$ for $K=2$ and 
different system sizes, $N=25$,$100$,$1000$,$10000$. One sees that 
the percolation transition is reached only for a too large size,
$N \sim 10000$ that is
far beyond the largest sample for which one can compute the
entropy by exhaustive enumeration. Moreover, it is 
important to notice that the approach to the asymptotic value 
is nonmonotonic since, for these values of $N$ 
the approach to the asympote comes from below while one can easily 
prove that $P_{\sc max}(N=4,\alpha=0.75)\sim 0.75$.

Finally, the third curve in Fig.~\ref{entropy_max} (stars) represents 
the ``improved'' contribution of the 
largest cluster, 
$0.5828 \; \overline{S_{\sc max}/L_{\sc max}}$. For $N>20$ 
the improved curve is still 
higher than the actual one and a linear fit yields the
limiting value $\lim_{N\to\infty} \overline{S_{\sc max}}/N \sim
0.2143 \pm 0.0003$.

This numerical analysis, even if based on simulations of very small 
systems and highly speculative, 
suggests that the contribution of the percolating cluster
is finite in the thermodyamic limit.  
On the analytical side, it might be possible to compute the 
entropy of the percolating cluster, 
at least for  2-sat, by 
taking profit of the numerous mathematical results on the
structure of random graphs~\cite{Bollobas}. Such a study is necessary
to settle the problem. 

This picture could also explain the discrepancy between numerical
studies and a rigorous bound for the value of the exponent governing
the size of the scaling window of the satisfiability
transition~\cite{Wilson,scaling2}. For the  relatively small system sizes
first studied, the observed exponent may have been altered by
percolation, as its asymptotic regime was not reached yet~\cite{Remi2}. 
At larger sizes, and in
mathematical studies, the exponent is purely due to the satisfiability
threshold inside the giant cluster; one might observe the true
exponent for lower system sizes by studying the probability of the
largest cluster to be satisfiable.

Let us summarize briefly the route
followed in studies based on the replica trick
to attempt to identify the effect of the 
percolation transition within this calculation~\cite{Remi}. In the {\sc
rs} {\it Ansatz}, an intricate integral equation over the probability of 
local fields, 
$P(h)$, that is the order parameter of the system, is obtained. From 
$P(h)$ all thermodynamic quantities follow,
including the entropy of the satisfiable phase. As the analytical
resolution of the equation that determines $P(h)$ seems out of reach, it has
been solved order by order in $\alpha$. This yields an $\alpha$
expansion for the entropy which is in exact coincidence with ours up
to $O(\alpha^4)$ for $K=2$ and up to $O(\alpha^2)$ 
for $K \ge 3$. It seems natural to
assume that the two expansions coincide to all orders. The {\sc
rs} {\it Ansatz} is thus proven to be exact for all $\alpha$ 
such that 
$\alpha< \alpha_p=1/(K(K-1))$. Beyond this value
if, as we discussed above, the influence of the loops on  
the entropy of the percolating cluster is not negligible,
two possibilities arise:
either the {\sc rs} {\it Ansatz} is
wrong, or a more refined handling of the integral equation on $P(h)$
is required~\cite{comment}. A careful analysis of this problem is 
worthwhile. One could, for instance, investigate the presence of a
singularity at the percolation threshold. In any case, the fact that
the entropy of the satisfiable phase remains finite up to the
satisfiability transition is confirmed: at the threshold a finite,
even if small ($\sim 0.2$ for $K=2$), 
fraction of sites are in finite size clusters, and
their contribution provides a lower bound for the entropy of the system.

\section{Dilute spin glasses: an exact solution for the
paramagnetic phase of the Viana Bray model}
\label{vbsection}

Spin glasses are magnetic systems where the interactions between
degrees of freedom (spins) are disordered. The key quantity that determines 
all the statistical properties of 
such systems is the free energy averaged over the
distribution of the interactions. In most cases 
this average can
only be computed with the replica trick~\cite{beyond}, often involving
the technically subtle replica symmetry breaking ({\sc rsb}). For infinitely
connected models, in which each spin interacts with all others
in the sample,
like the Sherrington-Kirkpatrick model~\cite{SK},
the {\sc rsb} scheme that yields the 
exact solution in both high and low-temperature phases 
is well understood. The final aim, far
from being reached, is to determine the nature of the 
spin-glass phase of disordered models on a
$d$-dimensional lattice with only short range
interactions. Dilute spin glasses can be viewed as an intermediate
step between these two limits: each spin interacts with a finite
number of other spins but the model includes no notion of distance
since the ``neighbors'' are randomly chosen from the whole set of 
spins of the system.

The standard dilute spin-glass model has been introduced by Viana and
Bray~\cite{viana}. It is defined by the following  
Hamiltonian involving $N$ classical
Ising spins, $S_i=\pm 1$,
\begin{equation}
H=-\sum_{i\neq j} J_{ij} S_i S_j
\; .
\end{equation}
The interactions $J_{ij}$ are independently distributed with the
same probability law:
\begin{equation}
P(J_{ij})=\left(1-\frac{c}{N}\right)\delta(J_{ij})
+\frac{c}{N}\rho(J_{ij})
\; .
\end{equation}
$\rho$ is normalized to one, and has average and mean square deviation of
order one to obtain a sensible thermodynamic limit. $c$ is the mean
connectivity per spin.

Despite numerous studies~\cite{viana,diluted,other_diluted}, 
a complete understanding of this model has not been 
reached yet. The main difficulty in the study of dilute models is
that even without {\sc rsb}, the order parameter is a function instead of a
number as in the {\sc ic} case. In order 
to introduce {\sc rsb} one has to cope with an
order parameter which is at least a functional, leading
to very difficult calculations.

The mean connectivity per spin 
$c$ plays here the same role as $\alpha$ in our expansion of the 2-sat
entropy. As opposed to K-sat, for the Viana-Bray model the percolation and 
the paramagnetic ({\sc pm}) to spin-glass transition occur at the same critical
value $c=1$. In the dilute phase, $c<1$, 
the clusters have not percolated and the model is in the {\sc pm} 
phase~\cite{KS}. The statistical properties in this  phase 
can be studied with the cluster expansion. The average
free-energy per site reads
\begin{equation}
-\beta f = 
\sum_t \frac{1}{L_t} c^{L_t-1} e^{-L_t c} K_t \; \overline{\ln
Z_t}
\; ,
\end{equation}
where the over-line  denotes an average with respect to
$\rho$ and only tree clusters contribute in the thermodynamic
limit. This is the analog of Eq.~(\ref{new_eq}) with a slightly different
expression for $P_t$ where $2\alpha$ is replaced by $c$.
$Z_t$ is the partition function of the cluster.
One can easily prove that for trees,
\begin{equation}
\overline{\ln Z_t} = L_t \ln 2 + (L_t-1) \; \overline{\ln \cosh \beta
J}
\; ,
\label{fvbtree}
\end{equation}
irrespectively of the topology. Then the sum of symmetry factors for
all trees of $L_t$ sites is $L_t^{L_t-2}/(L_t-1)!$ (a well-known result
from graph theory \cite{Bollo2}), and results in 
\begin{equation}
- \beta f = \ln 2 
\left( \sum_{k=1}^{\infty}\frac{(e^{-c})^k}{k!}(ck)^{k-1}\right) 
+ \overline{\ln \cosh \beta J}
\left(c \sum_{k=1}^{\infty}\frac{(e^{-c})^k}{k!}(k-1)(ck)^{k-2}\right)
\; .
\label{sumVB}
\end{equation}
The two sums can be evaluated (cf Appendix) to yield
\begin{equation}
-\beta f = \ln 2 +\frac{c}{2} \; \overline{\ln \cosh \beta J}
\; .
\end{equation}
This result has a simple interpretation: from Eq.~(\ref{fvbtree}), 
each site contributes with $\ln 2$ to the free energy,
each link with $\overline{\ln \cosh \beta J}$ for tree-like
sructures. As there are $N$ sites and $cN/2$ links on average, the
result follows. One obtains exactly the same free energy following the
replica method with a {\sc rs} {\it Ansatz}. The exactness of the {\sc
rs} {\it Ansatz} in this phase is then proven. 

The cluster expansion allows to
compute finite size corrections in a rather simple manner. These 
read, upto $O(c^3/N)$,  
\begin{eqnarray*}
\frac{1}{N}\left[ -\frac{c}{2}\; \overline{\ln \cosh \beta J} + c^3
\left( \frac{4}{3} \ln 2 + \overline{\ln \cosh \beta J} + \frac{1}{6}
\; \overline{ \ln (1+\tanh \beta J_1 \tanh \beta J_2 \tanh \beta J_3 ) }
\right) \right]
\end{eqnarray*}
where $J_1$, $J_2$ and $J_3$ are three independent couplings taken from
the probability distribution $\rho(J_{ij})$.
It will be very interesting to confront this result with the 
finite size corrections to the replica calculation 
using the {\sc rs} {\it Ansatz}.

As in the K-sat problem, 
the expansion cannot be used beyond $c=1$, since it does not
take into account the giant cluster appearing at the percolation
transition.

\section{Conclusions and perspectives}
\label{concl}

The cluster expansion  relies on very simple
combinatorial arguments. It allows one to solve 
optimization problems in the ``easy'' phase 
avoiding the introduction of replicas and  
it is a general method to obtain finite $N$ corrections.
Most importantly, it has allowed us to signal 
the possible need for a revision of the replica 
solution of the K-sat, and similar problems, with two 
successive percolation and easy-to-hard transition.

For spin-glasses without a difference in these two transitions
the interest of the method is less apparent, as highly diluted
systems are in the less interesting 
{\sc pm} phase. Still, it would 
be interesting to test if the {\sc rs} {\it Ansatz} 
is exact for any spin-glass model under 
its percolation threshold, as was proven
here for the Viana-Bray spin-glass in the thermodynamic limit.

Let us note, however,  the difference in the percolating and 
critical behavior of K-sat and the Viana-Bray
dilute spin-glass. In the former, percolation occurs before 
the satisfiability transition ($\alpha_p < \alpha_c$);
in the latter both phenomena arise 
at the same value of the control parameter
($c=1$). This can be understood as being due to the fact that 
2-sat is, in a way,  less frustrated than VB. 
Two manifestations of this fact are given by 
the behavior of a single loop and a linear cluster. 
A single loop in 2-sat is always satisfiable while in VB 
it is frustrated each time there is an odd number of antiferromagnetic 
couplings on it. A linear cluster in 2-sat can be satisfied by a 
large number of configurations while in VB only two spin configurations 
satisfy all bonds. 

The cluster expansion can be applied to a variety of interesting problems.
For instance, 
algorithms that solve satisfiability problems through local search, like
walk-sat~\cite{walk-sat}, can also be studied with 
this method~\cite{new}. The number of
steps needed to solve a formula is the sum of the number of steps needed to
solve each cluster. Improvements of the algorithms by means of
better heuristics can thus be quantified~\cite{new}.

\bigskip
\noindent{\underline{Acknowledgements}}

Very useful discussions with S. Mertens,
R. Monasson and M. Weigt are gratefully acknowledged.  
GS warmly thanks D. S. Sherrington for introducing him to the 
subject of disordered systems and optimization problems during 
a long visit to Oxford. 
We acknowledge financial support from the ACI
``Algorithmes d'optimisation et syst{\`e}mes d{\'e}sordonn{\'e}s
quantiques''.  

\bigskip
\noindent{\large \bf Appendix}
\bigskip

\noindent 
In this appendix we shall derive  a proof of the two summations used in
eq. (\ref{sumVB}):

\begin{eqnarray}
A &=& \sum_{k=1}^{\infty}\frac{(e^{-c})^k}{k!}(ck)^{k-1} = 1 \; ,
\\
B&=&\sum_{k=1}^{\infty}\frac{(e^{-c})^k}{k!}(k-1)(ck)^{k-2}=\frac{1}{2} \; .
\end{eqnarray}
 
The proof relies on a
mathematical identity proven with the help of analytical function
tools~\cite{Arfken}. Let $w(z)$ be a given function which can be
inverted to $z(w)$. Then the coefficients of the serie expansion of
$z(w)$ are obtained via the following expression:

\begin{eqnarray}
z(w)&=&\sum_{k=1}^{\infty} \frac{1}{k!} b_k w^k
\; ,
\\
b_k &=& \frac{d^{k-1}}{dt^{k-1}}
\left[\left(\frac{t}{w(t)}\right)^k\right]_{t=0}
\; .
\end{eqnarray}

Let us consider the function $w(z)=z \exp(-c z)$, which is a bijection
from $[0,1]$ to $[0,e^{-c}]$ if $c<1$. The coefficients of the series
expansion of the reciprocal $z(w)$ are
\begin{equation}
b_k=\frac{d^{k-1}}{dt^{k-1}}
\left[\left(\frac{t}{te^{-c t}}\right)^k\right]_{t=0}=(kc)^{k-1}
\; .
\end{equation}
Thus, $A=z(e^{-c})=1$, as $w(1)=e^{-c}$.

The second series can be transformed using $A=1$ :

\begin{equation}
B=\frac{1-e^{-c}}{c}-\sum_{k=2}^{\infty}\frac{(e^{-c})^k}{k!} (ck)^{k-2}
\; .
\end{equation}
If we define
\begin{equation}
g(w) \equiv \sum_{k=2}^{\infty}\frac{w^k}{k!} (ck)^{k-2}
\; ,
\end{equation}
then  $B=(1-e^{-c})/c-g(e^{-c})$. To compute $g(w)$, we note that
$g(0)=0$ and $g'(w)=z(w)/(cw)-1/c$. By integration and with the change
of variables $z=z(w)$, one obtains
\begin{equation}
g(e^{-c})=-\frac{e^{-c}}{c} + \frac{1}{c} \int_0^1 dz \; w'(z)
\frac{z}{w(z)} = \frac{1-e^{-c}}{c}-\frac{1}{2}
\; .
\end{equation}
This yields the final result $B=1/2$.


\begin{thebibliography}{99}
\bibitem{comment} It is interesting to note that in the variational
calculation of Ref.~\cite{Biroli}, the replica symmetry breaking 
solution ({\sc sc rsb}) of $K\leq 3$-sat appears at
a value $\alpha_{\sc rsb}$ of $\alpha$ 
in between $\alpha_p$ and $\alpha_c$. Due to the 
variational character of the solution in~\cite{Biroli}, 
the $\alpha_{\sc rsb}$ is an upper bound 
for the exact {\sc rsb} threshold. If the appearence of the {\sc rsb} 
solution is related to some percolation phenomenon remains 
an interesting open issue.
\end{thebibliography}
\end{document}